\documentclass[11pt]{article}
\usepackage{appb,epsfig}

\begin{document}
\title{
Coulomb Effects in Charged Particle Spectra\\ 
from Heavy Ion Collisions at AGS and SIS
\thanks{Presented at MESON'98, Cracow, Poland, May 30 - June 2, 1998}
}
\author{C.~M\"untz
\address{Univ.~of Frankfurt, for the E802/E866 and KaoS Collaborations}
}
\maketitle
\begin{abstract}
Data from the AGS experiment E866 on charged particles emitted
in central Au+Au collisions 
at 10.8 AGeV incident energy is reviewed.
The study of spectral shapes indicates the presence of 
collective transverse
flow and -- in particular for charged pions -- 
of Coulomb final state interaction
between emitted particles and the nuclear charge distribution. 
The comparison to SIS pion data measured with KaoS elucidates
the role of charged pions to probe nuclear reaction dynamics
at lower incident energies.
\end{abstract}
  
\vspace*{-0.3cm}
\section{Experimental results at 10.8 AGeV incident energy}
\vspace*{-0.2cm}
Experiment 866, installed at AGS/BNL, was designed to
investigate Au+Au reactions in the energy regime of 10~AGeV~\cite{e866spec}. 
The E866 spectrometers provide particle identification
and momentum measurement for a variety of charged particles. 
Central reactions can be selected by means of an event characterization.
{\small
\begin{figure}[ht]
\begin{center}
\begin{minipage}[h]{12.6cm}
\begin{center}
\mbox{\epsfig{width=12.5cm,height=7.5cm,
file=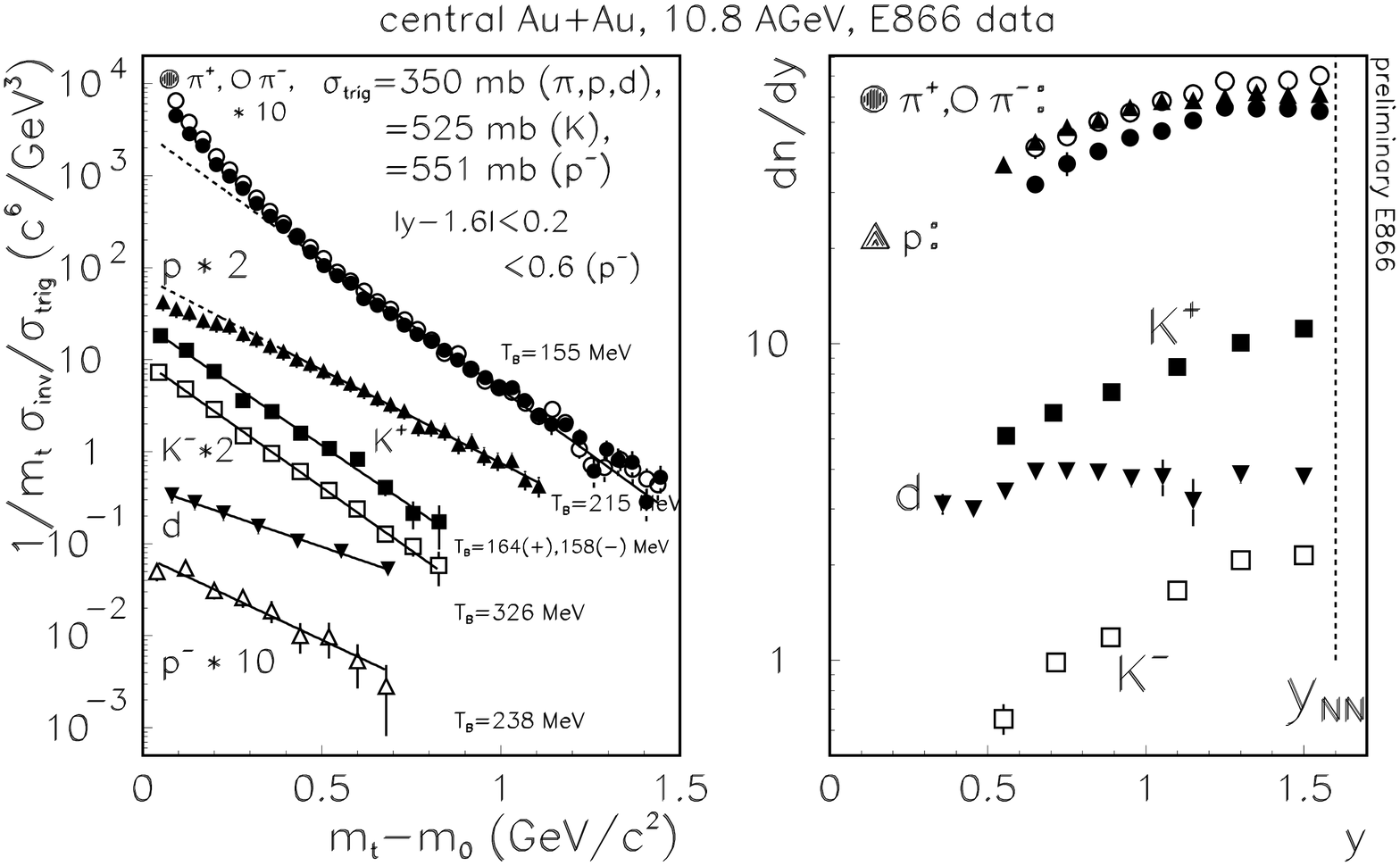
}}
\vspace*{-0.6cm}
\caption{
{\it
Data from E866. Left: Boltzmann representation of spectra
measured around midrapidity in central Au+Au collisions. $T_B$ denotes
the inverse slope parameter. Right: rapidity distributions, shown in one 
hemisphere. Besides pion and proton distributions data is preliminary.
Statistical errors only, systematic errors estimated to be $\pm(10-15)$\%.
}}\label{spectra}
\end{center}
\end{minipage}
\end{center}
\vspace*{-0.9cm}
\end{figure}
}
Figure~\ref{spectra} summarizes recent, partly
preliminary results studying central
Au+Au collision at 10.8~AGeV bombarding energy~\cite{e866data}.
The rapidity distributions depicted on the right panel demonstrate the large 
phase space around midrapidity covered by the experiment. 
The distributions of protons and charged mesons are peaked at midrapidity,
deuterons exhibit a rather flat distribution. The protons pile up
at midrapidity confirms the expectation of a large amount of stopping and the
concomitant expectation of high baryon density in central Au+Au collisions.

Corresponding m${}_t$ spectra of particles 
emitted at midrapidity are plotted on the
left panel in a way that thermal spectra would appear as a straight line. 
Studying the inverse 
slope parameters~$T_B$ of the spectral distributions an almost 
linear increase of $T_B$ 
with the particle mass results. 
The inverse slope parameters of the heavier particles can hardly
be described within a thermal picture alone. 
Similar observations have been made at SPS energies~\cite{sps}. 

The mass dependence of $T_B$ suggests taking into account flow effects. 
Accordingly,
the data presented here have been analyzed in transverse direction assuming
an exploding fireball scenario~\cite{hirsch97}. A common 
freeze-out temperature of $127^{+10}_{-15}$~MeV and an average transverse flow 
velocity $<\beta_t>$=0.39$\pm$0.05 results which
fits to a trend given by similar
analysis in the SIS/BEVALAC and SPS energy regimes.

Most recently, preliminary data 
on charged particle production measured by E866 and E917
in the energy range between 1 and 10 AGeV has become available~\cite{e917}.
Their analysis will systematically complete the gap in incident energy 
between SIS/BEVALAC and AGS.

\vspace*{-0.3cm}
\section{Charged particle ratios and Coulomb effects} 
\vspace*{-0.2cm}
Positively and negatively charged pions exhibit a significant 
difference at low m${}_t$, already visible in the compressed scale of
fig.~\ref{spectra}, left panel, in addition to their common low-m${}_t$
enhancement. Figure~\ref{ratios} studies this difference in detail, 
{\small
\begin{figure}[ht]
\begin{center}
\begin{minipage}[h]{12.6cm}
\begin{center}
\mbox{\epsfig{width=7.5cm,height=7.5cm,
file=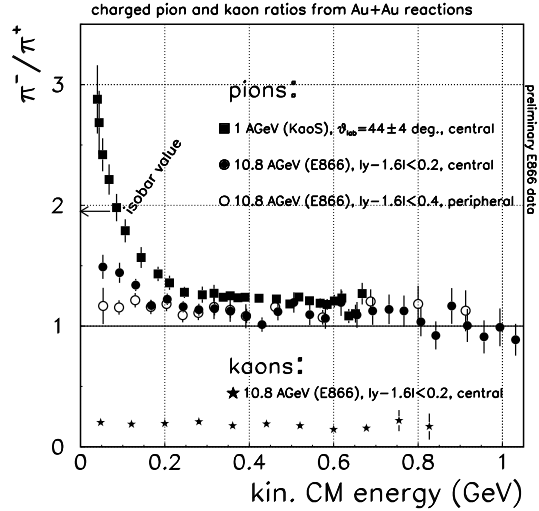
}}
\vspace*{-0.2cm}
\caption{
{\it 
Energy differential ratios of spectral distribution of positively and
negatively charged pions and kaons 
at 1 and 10.8 AGeV incident energy~[6,2]. 
Preliminary data, besides the pion data at central reactions.
Statistical errors only.
}}\label{ratios}
\end{center}
\end{minipage}
\end{center}
\vspace*{-0.9cm}
\end{figure}
}
and for different reactions, including all 
available data~\cite{e866data,kaos} at
1 and 10.8 AGeV incident energy. 
The following observations can be made: 
(i) The $\pi^-/\pi^+$ ratio rises at low pion CM energies for central Au+Au 
collisions both at 1 and 10.8~AGeV~\footnote{Hyperon decays can be excluded 
causing this effect at 10.8~AGeV.}.
(ii) The high-energy asymptotic value of this ratio, beyond 400 MeV, levels
off close to one.
(iii) The ratio of pions from peripheral reactions at 10.8 AGeV does
not depend on the pion energy. It agrees with the integrated negative to
positive pion ratio of 1.20$\pm$0.15 measured at central collisions.
(iv) The ratio of charged kaons does not depend on the kaon CM energy.
And most important,
(v) the low-energy rise of the pion ratio is most pronounced 
at 1~AGeV, exceeding the isobar value of 1.95 which
reflects the N/Z asymmetry of the colliding system. It is 
intriguing to note that the measured integrated negative to positive
pion ratio is 1.9$\pm$0.1 at this incident energy. 

These observations point to the presence of significant Coulomb
final state interactions between emitted charged pions and the remaining
nuclear charge distribution.
Hence, data are analyzed in the framework of a static Coulomb source resting
at midrapidity~\cite{coul}, giving rise to a 
distortion of spectral distributions due to both 
a momentum shift and a change of the momentum phase-space density.
The assumption of a static scenario
is motivated by considering pions with kinetic CM energies
above 60~MeV only. The corresponding CM velocities ($\beta_{CM}>$0.7) are 
significantly larger than the expansion velocity of the nuclear 
fireball ($\beta_t<$0.45~\cite{hirsch97}).

At AGS energies the pion spectra and the corresponding ratio can be
well described within this static scenario, yielding a rather moderate
Coulomb energy of 9$\pm$3~MeV. 
The corresponding Coulomb potential
does not affect the charged kaons
within the experimental acceptance. 
Similar results consistent with a moderate
effective Coulomb potential have been reported for SPS data~\cite{na44}.
In contrast, using the same static approach to describe the SIS data fails. 
An detailed analysis~\cite{kaos} including pion energy 
dependent Coulomb energies suggests, that low-energy pions feel rather
moderate Coulomb forces at freeze-out, while 
high-energy pions, which are produced
well below nucleon-nucleon threshold at 1~AGeV incident energy, are exposed
to a rather high Coulomb potential above 20~MeV when freezing-out.

In Reference~\cite{kaos} the variation of the Coulomb energy with the 
pion CM energy is interpreted as an experimental evidence for different
freeze-out radii of high- and low-energy pions emitted in central Au+Au
collisions at 1~AGeV: low-energy pions are emitted predominantly from
a more dilute nuclear charge distribution with consequently larger effective
freeze-out radii compared to high-energy pions, emitted from a more compact
source.

Refined model calculations~\cite{theory} 
support this interpretation and suggest that
charged pion spectra probe the dynamics of nuclear expansion in
relativistic heavy ion collisions. In particular at low
incident energies, different 
freeze-out radii can be linked to pion emission during different stages
of the reaction. Whereas low-energy pions are emitted later during expansion,
high-energy pions freeze-out early~\cite{mue}, still at a time where the 
nuclear density is enhanced and multiple baryon-baryon collisions
preferentially trigger the production of energetic subthreshold 
pions and kaons. 
In contrast, the comparably weak signal of charged pion ratios 
at higher bombarding energies indicates, that
the Coulomb source disintegrates fast
compared to the time scale of pion emission. Consequently, 
the extracted Coulomb 
energy represents an averaging over reaction dynamics.

\end{document}